# Hidden universality in large-deformation soft contact

Tong Mu [1,2,3], Changhong Linghu [3,*], Ruozhang Li [2,3,4], Shizhuo Weng [2], Jingyi Yu [2,3,5], Yanju Liu [6], Jinsong Leng [1,*], K. Jimmy Hsia [2,7,8,*], Huajian Gao [9,*]

1. *Centre of Composite Materials and Structures, Harbin Institute of Technology (HIT), Harbin 150080, China*
2. *School of Mechanical and Aerospace Engineering, Nanyang Technological University, 50 Nanyang Avenue, Singapore 639798, Singapore*
3. *Department of Mechanical Engineering, College of Engineering, City University of Hong Kong, 999077, Hong Kong, China.*
4. *State Key Laboratory of Mechanical System and Vibration, School of Mechanical Engineering, Shanghai Jiao Tong University, 800 Dongchuan Road, Shanghai 200240, China*
5. *School of Mechanical Engineering, Zhejiang University, Hangzhou, 310027, China*
6. *Department of Astronautical Science and Mechanics, Harbin Institute of Technology (HIT), Harbin 150080, China*
7. *School of Chemistry, Chemical Engineering and Biotechnology, Nanyang Technological University, 50 Nanyang Avenue, Singapore 639798, Singapore*
8. *School of Mechanics and Mechanical Engineering, Eastern University of Science and Technology, 2911 Haijiang Avenue, Ningbo 315200, China*
9. *Mechano-X Institute, Applied Mechanics Laboratory, Department of Engineering Mechanics, Tsinghua University, Beijing 100084, China*

[*] *Email: clinghu@cityu.edu.hk*

[*] *Email: lengjs@hit.edu.cn*

[*] *Email: kjhsia@ntu.edu.sg*

[*] *Email: gao.huajian@mail.tsinghua.edu.cn*

## Abstract

Large-deformation soft contact mediates how biological and robotic systems interact with their environments, underpinning embodied intelligence, wearables, human-machine interactions and tactile sensing. The same coupling between contact area and remaining thickness shapes the response of physical interaction interfaces in wearable sensors and soft robots. Yet predictive theories remain confined to small strains or specific geometries, and a general law for nonlinear soft contact has remained elusive. Here we reveal hidden universality in large-deformation soft-body contact: after normalization by a simple geometric correction, diverse force–displacement responses collapse onto a master curve. By extending the method of dimensionality reduction into the nonlinear regime, the framework yields analytical solutions for contact force and radius. Finite element analyses and experiments across canonical geometries, non-axisymmetric structures and irregular gummy candies validate the scaling law up to 50% compression. These results establish a universal principle for nonlinear soft contact and a mechanics-guided route to wearable devices, tactile systems and soft robotics interactions beyond empirical calibration.

# 1. Introduction

Large-deformation soft contact mediates how biological and robotic systems interact with their environments, underpinning embodied intelligence (1-3) , wearable interfaces (4-9), human-machine interactions and tactile sensing (10-18), and soft manipulation (19-24) . As illustrated in Fig. 1A, such contact spans multiple scales—from micrometer-scale deformation of cells and tissues (25-36) to macroscopic applications in flexible electronics (4-9, 37-41), smart adhesives (19-24, 42), meta-interfaces (43), and soft robotics (1-3). The geometries involved are equally diverse, including spheres, cones, cylinders, and irregular or anisotropic forms. This geometric diversity creates an apparent complexity: different shapes exhibit distinct contact responses, especially when deformation becomes large.

Despite extensive study—addressing geometry (44-46), adhesion (47-54), nonlinear elasticity (55-57), viscoelasticity (58, 59), and plasticity (60-62)—most theories are rooted in infinitesimal deformation and are therefore restricted to small strains. Sneddon (44) derived analytical solutions for axisymmetric contacts, and subsequent studies established power-law force–displacement relations (45, 46) for canonical geometries under linear elasticity (Fig. 1B). Yet extending such analytical descriptions into the nonlinear, large-deformation regimes remains challenging. Existing models are typically geometry-specific (55, 56) or semi-empirical (63, 64). Recent closed-form solutions for spherical compression under extreme compression have advanced this direction (57), but a general framework that unifies diverse geometries and materials (Figs. 1C-D) remains elusive.

Here we reveal a hidden universality in large-deformation soft-body contact. We extend the method of dimensionality reduction to derive the force and contact radius for axisymmetric power-law profiles. After accounting for the small-displacement response of each shape, we find that a common geometric correction captures the additional force required at finite compression. Experiments and finite element simulations test this scaling across canonical geometries, non-axisymmetric structures and irregular soft objects. At a physical interaction interface, the model relates the profile and thickness of a compliant tip to the force it will transmit at a prescribed displacement. This gives designers a way to choose those dimensions for a target load. The resulting scaling law organizes large-deformation soft contact around a small set of mechanical variables and reveals a shared geometric origin of nonlinear stiffening.

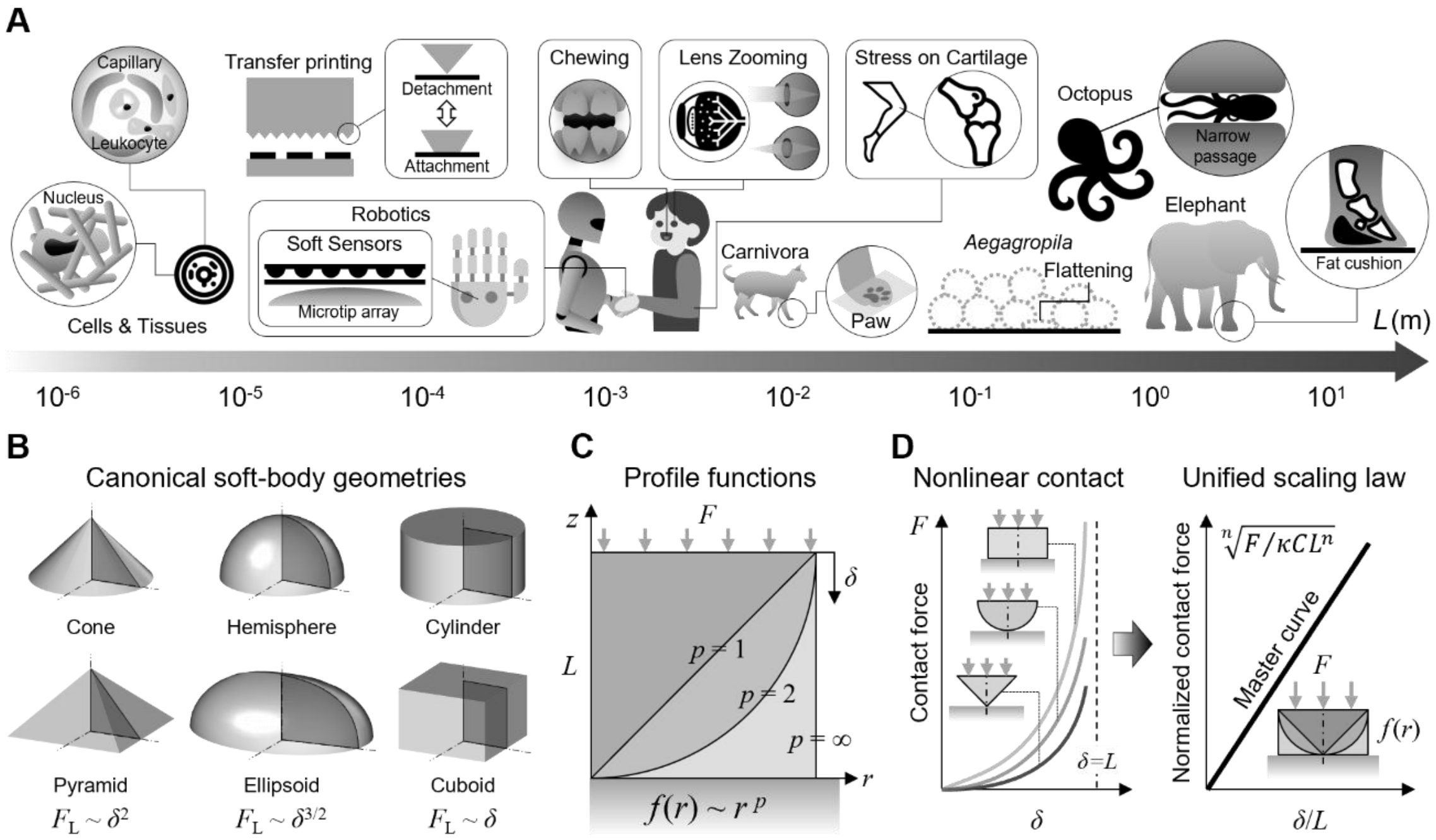

**Fig. 1. Large-deformation soft contact across biological and engineered systems and its hidden universality.** (**A**) Examples range from cells (nuclear deformation in confined environments (25, 26), leukocytes squeezing through capillaries (27, 28)) and humans (chewing food (32), accommodation of the crystalline lens (29, 30), articular cartilage compression (31)) to animals (paw–ground contact in carnivores (33), fat-cushion treading in elephants (34), octopus passing through narrow gaps (35)), plants (packing of *aegagropila* algae (36)), and engineering devices (microtip-array pressure sensors (4-6) and transfer-printing stamps (19-22)). (**B**) Canonical soft-body geometries (cone, hemisphere, cylinder, pyramid, ellipsoid, cuboid) idealize these diverse soft contact processes and follow distinct power-law force–displacement relations under small deformations. (**C**) Profile functions and compression schematics of these representative geometries. (**D**) Hidden universality in large-deformation soft contact. After normalization with a correction function $\kappa$, force–displacement responses from diverse geometries collapse onto a single master curve, revealing a general scaling principle.

# 2. Results and Discussion

## 2.1. Nonlinear method of dimensionality reduction for soft-body compression

Figure 2A illustrates the theoretical framework for modelling large-deformation compression of soft bodies. The system consists of a three-dimensional elastic body compressed against a rigid substrate by a load applied to its flat, constrained top surface, while permitting lateral expansion. For axisymmetric body geometries defined by a surface profile $f(r)$, the complex 3D contact problem can be reduced to a one-dimensional equivalent model. This approach generalizes Popov and Hess's method of dimensionality reduction (MDR, see *SI Appendix*, Sections S1) (46) from linear contact to the large-deformation regime. The equivalence guarantees that, for a given indentation depth $\delta$, the reduced model reproduces both the contact force $F$ and the contact radius $a$ of the full 3D system.

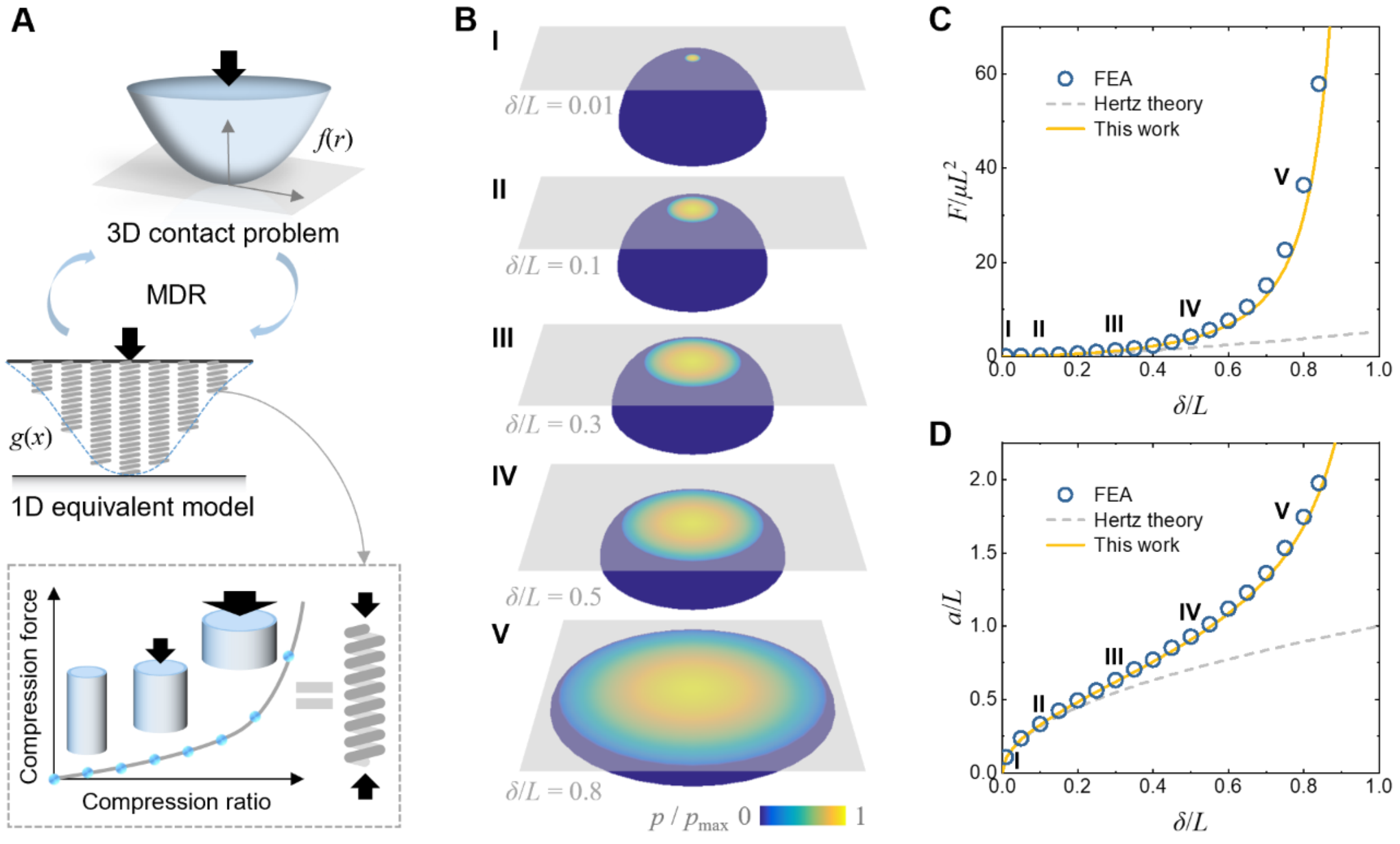

**Fig. 2. Modeling of soft-body compression under large deformations.** (**A**) Mapping of the 3D contact problem with axisymmetric profile $f(r)$ to a 1D equivalent model via the method of dimensionality reduction (MDR). The 1D system consists of nonlinear springs of varying lengths, reproducing the same contact force and radius as the 3D system under equivalent compression. Each spring corresponds to a cylinder of the same hyperelastic material. (**B**) Finite element simulations showing geometric nonlinearity in hemisphere compression, with progressive flattening at $\delta/L = 0$, 0.1, 0.3, 0.5, and 0.8. (**C**, **D**) Comparison of contact force and contact radius predicted by Hertz theory (gray dashed), the proposed model (yellow solid), and FEA (blue circles). Hertz theory deviates substantially at large deformations, whereas the proposed model incorporating geometric nonlinearity closely matches FEA.

As shown in Fig. 2B, FEA simulations reveal pronounced flattening at large compression ratios ($\delta/L \geq 0.3$), indicating substantial geometric nonlinearity. Figures 2C-D compare predictions from this framework with FEA and classical Hertz theory. Hertz theory (gray dashed lines) severely underestimates both contact force and radius once $\delta/L$ exceeds 0.3, underscoring the need to account for geometric nonlinearity. In the nonlinear MDR model, this is achieved by introducing nonlinear springs that account for finite-thickness and radial-expansion effects. Accordingly, the 1D equivalent model (Fig. 2A) consists of a continuum of nonlinear springs that reproduce the compression response of cylinders made of the same hyperelastic material, with the 1D profile $g(x)$ analytically related to the 3D geometry (derivations in *SI Appendix*, Sections S2). Additionally, as long as the constitutive models of the material share the same linear elastic regime, their influence on the compression response is minor (*SI Appendix*, Section S3). Thus, the neo-Hookean model is adopted as a representative in subsequent analyses.

For a representative power-law profile $f(r) = c\,r^p$, the linear solution gives the contact force as:

$$F_\mathrm{L} = C \cdot \delta^n, \qquad n = 1 + \frac{1}{p}, \tag{1}$$

where the scale factor $C$ is dependent on the material and geometry (see *SI Appendix*, Section S4).

The contact force is obtained from the nonlinear MDR in closed form as:

$$F = \kappa_n\left(\frac{\delta}{L}\right) \cdot F_{\mathrm{L}}, \tag{2}$$

where the nonlinear correction function is given by:

$$\kappa_n\left(\frac{\delta}{L}\right) = \frac{1 - \dfrac{2n}{1+n}\dfrac{\delta}{L} + \dfrac{n}{2+n}\left(\dfrac{\delta}{L}\right)^2}{\left(1 - \dfrac{\delta}{L}\right)^2}. \tag{3}$$

Likewise, the contact radius of power-law profile is calculated by the nonlinear MDR as:

$$a = \frac{1}{p} B_{\delta/L}\left(\frac{1}{p}, \frac{3}{2}\right) \cdot \frac{(\delta/L)^{-1/p}}{\sqrt{1-\delta/L}} \cdot a_{\mathrm{L}}, \tag{4}$$

where $B_x(a, b)$ denotes the incomplete beta function (65) (see solutions for some typical profiles in *SI Appendix*, Table S1), and the linear solution is:

$$a_{\mathrm{L}} = D \cdot \delta^{1/p}, \tag{5}$$

with the scale factor $D$ determined by the geometry analytically (see details in *SI Appendix*, Section S4).

### 2.2. Experimental validation of the universal theoretical framework

To validate the proposed nonlinear MDR theoretical model for large-compression contact of soft bodies, compression experiments (Fig. 3A) were conducted on soft structures with three representative geometries: cone, hemisphere (asymptotically expanding into a parabola; see *SI Appendix*, Section S4), and cylinder, corresponding to power-law profiles with exponents $p = 1$, $p = 2$, and $p = \infty$, respectively. Each sample was affixed to a rigid glass cylinder connected to the load cell of a universal testing machine and compressed against a fixed glass sheet. A digital camera was positioned laterally to record the evolution of the contact configuration during loading.

The selected geometries (Fig. 3B) span the representative range of shape profiles from sharp to flat, enabling evaluation of the model across a broad spectrum of curvature conditions. Side-view images of the samples during loading (Fig. 3C) illustrate the increasing contact area with compression displacement. Quantitative comparisons between model predictions and experiments are presented in Figs. 3D-E. Specifically, the contact force and contact radius were measured as functions of the normalized compression displacement $\delta/L$, and compared against the predictions of both the proposed model (solid lines) and linear solutions (dashed lines).

The results show very good agreement between the proposed model and experimental data across all geometries examined, while the linear predictions deviate significantly, particularly under large deformations. This confirms that our model effectively captures the nonlinear compression behavior of soft bodies with various canonical geometries beyond the small-strain assumptions of classical theory.

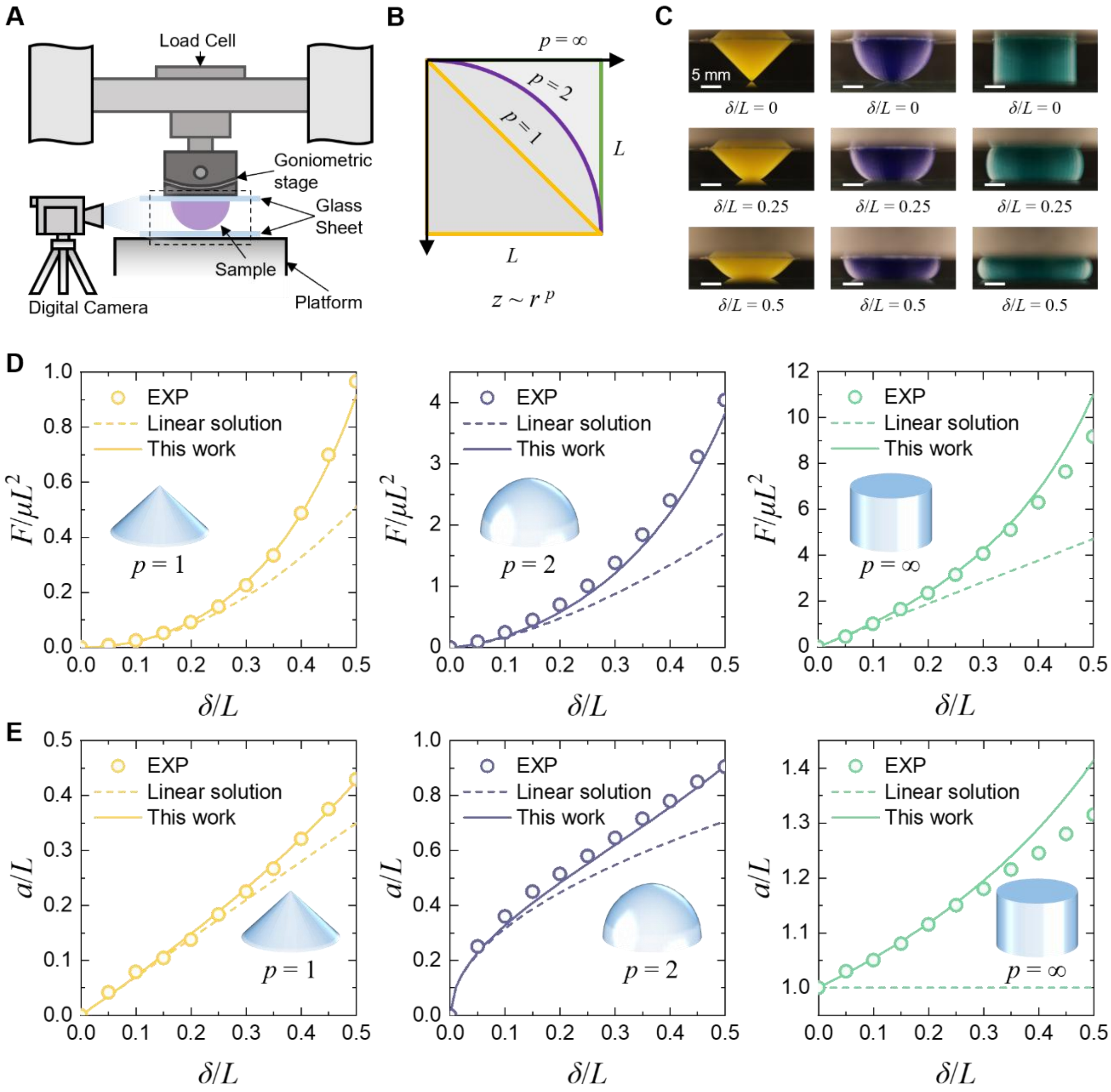

**Fig. 3. Experimental validation of the proposed compression model for canonical profiles.** (**A**) Experimental setup. Samples were mounted on a glass cylinder attached to a load cell and compressed against a fixed glass plate, while a lateral camera recorded the contact evolution. (**B**) Sample geometries described by power-law profiles: cone ($p$ = 1, yellow), hemisphere ($p$ = 2, purple), and cylinder ($p = \infty$, green), with characteristic length $L$ = 20 mm. (**C**) Optical images of samples at compression ratios $\delta/L$ = 0, 0.25, and 0.5. (**D**, **E**) Comparison of predicted (lines) and measured (circles) (**D**) contact force and (**E**) contact radius versus compression ratio. The proposed model (solid lines) agrees closely with experiments (circles) across geometries, while Hertz theory (dashed lines) deviates substantially at large deformations.

A slight deviation is observed in the case of the cylinder geometry, particularly at higher compressive strains. This discrepancy is likely due to frictional effects at the interface between the cylinder and the glass surface, as evidenced by the pronounced radial expansion of the sample at the contact interface in the experimental images (Fig. 3C). Even so, at a compression ratio of 0.5, the relative errors remain far smaller than those of Hertz theory: 20% in this model versus 48.6% in Hertz theory for contact force, and 7.5% versus 24% for contact radius. These comparisons further support the improved accuracy of the present framework.

### 2.3. Hidden universality across diverse geometric profiles

After experimental validation on canonical profiles of cone, hemisphere and cylinder, the nonlinear MDR framework was used to ask whether large-deformation soft-body compression contains a broader universality beyond these idealized shapes. Finite element analysis (FEA) simulations were conducted on additional geometries, including hemi-ellipsoid, pyramid and cuboid, together with the canonical profiles examined above (Fig. 4). These datasets span axisymmetric and non-axisymmetric cases and allow the geometry-dependent responses to be systematically normalized. The resulting scaling law was then examined experimentally using irregularly shaped soft candies, providing a practical validation beyond idealized geometries.

As shown in Fig. 4A, FEA results (scattered points) agree closely with theoretical predictions from the proposed model (solid lines) across six representative shapes: hemisphere, hemi-ellipsoid, cone, pyramid, cylinder, and cuboid. The agreement extends to non-axisymmetric geometries, suggesting that the global force response under large deformation is governed primarily by the evolution of effective contact area rather than rotational symmetry alone. Shapes with comparable profiles therefore exhibit similar confinement effects and nonlinear corrections. Accordingly, non-axisymmetric shapes can be described using the same correction function as their axisymmetric counterparts when they share the same exponent $n$ in Eq. (1), with parameter $C$ proportional to the base area.

To account for remaining geometric variations, shape-specific correction functions were calculated using Eq. (3) (inset of Fig. 4B). Despite differences among profiles, all correction functions fall within a narrow band, indicating that the influence of geometry is strongly constrained after normalization. This behavior is captured by a universal form (dashed black line), which provides an effective descriptor across disparate geometries. The universal correction function is expressed as:

$$\kappa\left(\frac{\delta}{L}\right) = \left(1 - k\frac{\delta}{L}\right)^{-1}, \tag{6}$$

where $k$ = 10/9 is a universal parameter obtained from correction functions across diverse shapes (as shown in the inset of Fig. 4B). Physically, $\kappa(\delta/L)$ represents a geometric stiffening factor arising from finite-thickness effects. As compression increases, the finite thickness of the soft body imposes a stronger constraint on the deformation field compared to a semi-infinite half-space, leading to systematic deviation from the classical linear scaling ($\kappa = 1$).

This finding suggests that the apparent shape dependence of contact force can be systematically encoded into a correction function that follows a predictable trend. When the contact force is normalized by the universal correction function, data from all geometries collapse onto a single master curve (Fig. 4B). The resulting universal scaling law of contact force for extreme compression is given by:

$$\frac{F}{\kappa} \sim \delta^{n}. \tag{7}$$

The revealed hidden universality simplifies the characterization of large-deformation contact in soft materials and enables force prediction from geometric parameters alone.

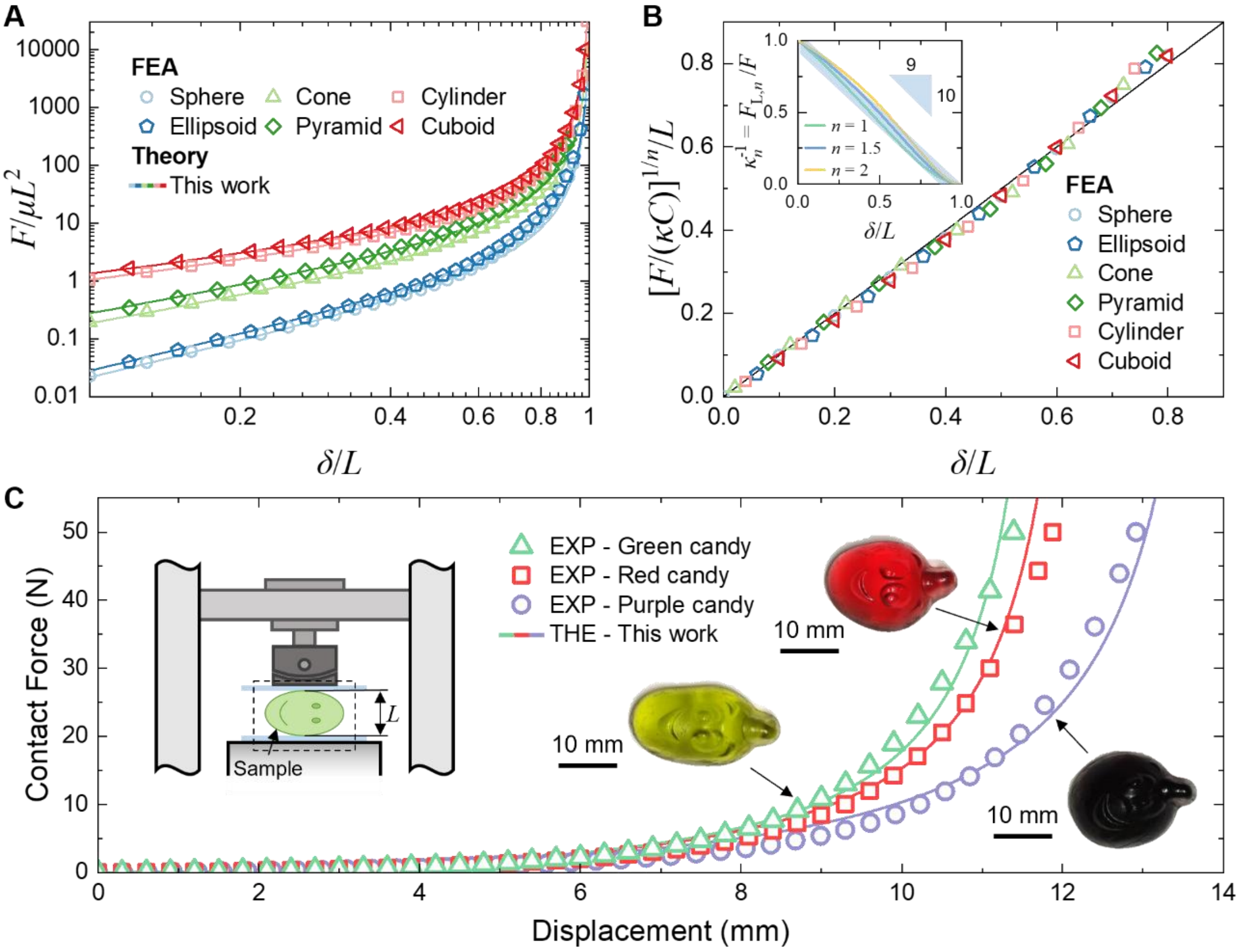

**Fig. 4. Hidden universality and universal scaling law for large-deformation compression of soft bodies.** (**A**) Contact force obtained from finite element analysis (FEA, scattered points) and predicted by the proposed model (solid lines) for representative geometries. (**B**) Normalized contact force for diverse geometries: hemisphere (light blue), hemi-ellipsoid (dark blue), cone (light green), pyramid (dark green), cylinder (pink), and cuboid (red). Inset: shape-specific correction functions for cylinder/cuboid ($n$ = 1, green), sphere/ellipsoid ($n$ = 1.5, blue), cone/pyramid ($n$ = 2, yellow), all falling within a narrow band (light blue). The universal scaling law (black dashed line) captures the common trend across shapes. (**C**) Compression of irregularly shaped soft candies. Experimental force–displacement data (green triangles, red squares and purple circles) align closely with predictions from the proposed scaling law (solid lines).

To further test the predictive power of the universal scaling law beyond idealized geometries, it was applied to commercially available gelatin-based gummy candies with irregular shapes (Happy Grapes, Haribo®). These samples were selected for their uniform composition, consistent manufacturing quality, and highly compliant mechanical characteristics, while providing distinct asymmetric geometries.

As shown in Fig. 4C, the three candies differ markedly from the canonical profiles considered above. Despite their geometric complexity, the scaling law successfully captures the compression behavior when an appropriate characteristic length $L$ is defined for each shape (green: $L$ = 13.5 mm; red: $L$ = 14.0 mm; purple: $L$ = 16.0 mm). Under small deformations, the response follows the Hertzian law, i.e., $F_L \sim \delta^{1.5}$. Under large deformations, the contact force can be predicted using the proposed scaling law according to Eq. (7). The close agreement between experiment (scattered points) and theory (solid lines) across all three cases confirms that the scaling law accurately captures the essential mechanics of large-deformation contact, even for irregular and non-axisymmetric soft objects.

## 3. Conclusion

This work reveals a hidden universality underlying large-deformation contact of soft bodies. By establishing a nonlinear dimensional reduction framework for soft-body compression against rigid substrates, analytical expressions for contact force and contact radius were derived and validated across multiple geometries and deformation regimes using experiments and finite element simulations. After normalization by a geometric correction, nonlinear force–displacement responses from canonical geometries, non-axisymmetric structures and irregular soft objects collapse onto a universal scaling law. These results advance the fundamental understanding of large-strain soft contact and provide a compact predictive principle for nonlinear contact mechanics.

Overall, this work shows how nonlinear soft-contact responses that appear geometry-dependent can be unified through a simple scaling law. This mechanics-guided principle provides a route to intelligent physical interfaces that are less dependent on empirical calibration, with implications for soft robotics, biomedical engineering and advanced manufacturing.

## 4. Materials and Methods

### 4.1. FEA setup

FEA simulations were conducted using the commercial finite element software ABAQUS to model the compressive behavior of soft bodies with various geometries (Figs. 2C-D and Figs. 4A-B). All soft structures were modeled as incompressible neo-Hookean solids with a shear modulus of $\mu$ = 1 MPa.

Cones, hemispheres and cylinders were represented by two-dimensional axisymmetric models with initial height and base radius of 1 mm. Four-node axisymmetric quadrilateral hybrid elements (CAX4H) were used, with a maximum element size below 0.2 mm. Pyramids, cuboids and hemi-ellipsoids were represented by three-dimensional models with initial height of 1 mm. The pyramid and cuboid had square bases with 2 mm edge length. The hemi-ellipsoid bottom face had reported semi-minor and semi-major axis dimensions of 1 and 1.5 mm, respectively. Eight-node linear brick hybrid elements (C3D8H) were used, with maximum edge length below 0.2 mm.

The substrate was modeled as an analytical rigid body with all degrees of freedom fully constrained. Surface-to-node contact was defined between the soft body and the rigid substrate using a frictionless, "hard" contact condition, which enforces a strict non-penetration constraint appropriate for the interaction between the soft elastomer and the rigid substrate. A static general step was used for all simulations, and geometric nonlinearity was enabled (NLGEOM = ON) to account for large deformations. A prescribed vertical displacement $\delta$ = 0.5 mm was applied to the top surface of the soft structures to induce compression. The normal contact force was extracted from the vertical reaction force (RF2) at the reference point of the substrate, and the contact radius was calculated based on the total contact area (CAREA) at the interface.

### 4.2. Sample preparation

Soft samples (Fig. 3) with designated geometries were fabricated using Ecoflex 00-30 (Smooth-On, Inc.), a platinum-cured silicone rubber widely used for soft mechanical testing due to its high deformability and near-incompressibility. The part A and part B were mixed thoroughly in a 1:1 weight ratio and thoroughly degassed under vacuum to remove trapped air bubbles. Custom molds with three typical shapes were fabricated with a 3D printer (Bambu Lab P1S) using PLA Basic filament. After curing at room temperature for 4 h, the silicone samples were demolded for compression tests.

### 4.3. Contact test

Compression tests were conducted on Ecoflex samples using a universal testing machine (Instron 5566A) equipped with either a 50 N load cell for cone and hemisphere samples or a 500 N load cell for cylinder samples. Each sample was affixed to the flat surface of a rigid glass cylinder using a thin layer of Vaseline petrolatum to ensure stable adhesion during loading. The samples were compressed vertically against a fixed transparent glass plate at a constant displacement rate of 0.1 mm/s, while the normal force was recorded continuously throughout the loading process.

To monitor the contact area evolution, a digital camera was placed laterally to capture images of the contact radius in real time. These images were analyzed using custom Python scripts to extract the contact radius at each displacement increment. All tests were conducted at room temperature, and three samples were tested in triplicate to ensure reproducibility.

Compression tests for Haribo soft candies (Fig. 4C) were conducted under the same conditions using the Instron 5566A with a 50 N load cell at a compression speed of 0.1 mm/s.

## Data, Materials, and Software Availability

All data are included in the article and/or SI Appendix.

## Acknowledgments

K.J.H., H.G., and C.L. acknowledge support from the Ministry of Education (MOE) of Singapore under the Academic Research Fund Tier 2 (MOE-T2EP50122-0001). Y.L. and J.L. acknowledge support from Fundamental and Interdisciplinary Disciplines Breakthrough Plan of the Ministry of Education of China (JYB2025XDXM413). C.L. acknowledges support by Hong Kong RGC Grant ECS (21214426) and the CityU Start-up Grant (9382018)., and the support from Key Laboratory of Low Dimensional Materials and Application Technology (Xiangtan University) of Ministry of Education (2025ZD03).

T.M. acknowledges support from the China Scholarship Council program (202406120073).

## Author contributions

T.M., C.L., R.L., Y.L., J.L., K.J.H., and H.G. designed research; T.M., C.L., S.W., J.Y., Y.L., J.L., K.J.H., and H.G. performed research; T.M., C.L., S.W., J. Y., Y.L., J.L., K.J.H., and H.G. analyzed data; T.M., C.L., S.W., J.Y., Y.L., J.L., K.J.H., and H.G. data visualization; and T.M., C.L., R.L., Y.L., J.L., K.J.H., and H.G. wrote the paper.

## Competing interests

The authors declare no competing interests.

# Supporting Information for

# Hidden universality in large-deformation soft contact

Tong Mu [1,2,3], Changhong Linghu [3,*], Ruozhang Li [2,3,4], Shizhuo Weng [2], Jingyi Yu [2,3,5], Yanju Liu [6], Jinsong Leng [1,*], K. Jimmy Hsia [2,7,*], Huajian Gao [8,*]

1. *Centre of Composite Materials and Structures, Harbin Institute of Technology (HIT), Harbin 150080, China*
2. *School of Mechanical and Aerospace Engineering, Nanyang Technological University, 50 Nanyang Avenue, Singapore 639798, Singapore*
3. *Department of Mechanical Engineering, College of Engineering, City University of Hong Kong, 999077, Hong Kong, China.*
4. *State Key Laboratory of Mechanical System and Vibration, School of Mechanical Engineering, Shanghai Jiao Tong University, 800 Dongchuan Road, Shanghai 200240, China*
5. *School of Mechanical Engineering, Zhejiang University, Hangzhou, 310027, China*
6. *Department of Astronautical Science and Mechanics, Harbin Institute of Technology (HIT), Harbin 150080, China*
7. *School of Chemistry, Chemical Engineering and Biotechnology, Nanyang Technological University, 50 Nanyang Avenue, Singapore 639798, Singapore*
8. *Mechano-X Institute, Applied Mechanics Laboratory, Department of Engineering Mechanics, Tsinghua University, Beijing 100084, China*

[#] *These authors contributed equally to this work.*

[*] *Email:* clinghu@cityu.edu.hk

[*] *Email:* lengjs@hit.edu.cn

[*] *Email:* kjhsia@ntu.edu.sg

[*] *Email:* gao.huajian@mail.tsinghua.edu.cn

## Section S1. Linear solution of compression problem

To provide theoretical context, we briefly summarize the classical linear-elastic solution for the contact problem with an axisymmetric profile, referred to here as the linear solution.

The axisymmetric contact problem was first analytically solved by Sneddon using the Hankel transform technique (1). Building upon this, Popov and Hess introduced the method of dimensionality reduction (MDR), a mathematically rigorous and computationally efficient approach that reformulates the original three-dimensional (3D) contact problem into a one-dimensional (1D) analog (2, 3). Owing to its geometric clarity and simplicity, MDR has since been widely adopted in contact mechanics (2-8).

In linear elastic theory, indentation and compression problems are equivalent due to symmetry and reversibility. As such, the solution obtained for indentation directly applies to compression of a soft elastomeric body with the same surface profile. As illustrated in Fig. S1, MDR replaces the original 3D contact configuration (Fig. S1A) with a simplified 1D model (Fig. S1B) consisting of a rigid indenter with transformed profile $g(x)$, pressing against a Winkler foundation, which is an array of independent linear springs with stiffness per unit length equal to $E^*$. The transformation from the original axisymmetric profile $f(r)$ to the equivalent 1D profile $g(x)$ is given by:

$$g(x) = |x| \int_0^{|x|} \frac{f'(r)}{\sqrt{x^2 - r^2}} dr \,. \tag{S1}$$

This transformation preserves critical physical quantities, such as the indentation depth $\delta$, contact force $F_L$, and contact radius $a_L$ (where $F_L$ and $a_L$ represent linear solutions). In particular, the indentation depth in the 1D model is related by (2) as:

$$\delta = g(a_L). \tag{S2}$$

To verify equivalence between the original and transformed models, the total contact force in the 3D problem can be expressed as:

$$F_L = \int_0^{F_L} d\tilde{F} = \int_0^{a_L} \frac{d\tilde{F}}{d\tilde{\delta}} \frac{d\tilde{\delta}}{d\tilde{a}_L} d\tilde{a}_L = \int_0^{a_L} \frac{d\tilde{F}}{d\tilde{\delta}} g'(\tilde{a}_L) d\tilde{a}_L \,, \tag{S3}$$

where the increment of contact force $dF$ can be obtained by integrating the increment of the contact pressure $dp_0$ as:

$$dF_L = \int_0^{a_L} dp_0(r) 2\pi r dr \,. \tag{S4}$$

According to reference (2), the increment of contact pressure d$p_0$ in the indentation problem under a small displacement increment of d$\delta$ is equal to the contact pressure distribution of a flat punch with radius $a_L$ on same substrate under same displacement d$\delta$ as:

$$dp_0^{inf} = \frac{E^*}{\pi} \frac{1}{\sqrt{a_L^2 - r^2}} d\delta, \qquad (0 \leq r \leq a_L). \tag{S5}$$

The corresponding force increment is:

$$dF_L = \int_0^{a_L} dp_0^{inf} 2\pi r dr = 2E^* a_L d\delta. \tag{S6}$$

Thus, the contact force is:

$$F_L = 2E^* \int_0^{\delta} a_L d\tilde{\delta}\,. \tag{S7}$$

Integrating by parts, the contact force becomes:

$$F_L = 2E^* a_L g(a_L) - \int_0^{a_L} g(\tilde{a}_L) \frac{d}{d\tilde{a}_L} \frac{d\tilde{F}_L}{d\tilde{\delta}} d\tilde{a}_L\,. \tag{S8}$$

Substituting Eq. (S6) into Eq. (S8) and rearranging, it is found:

$$F_L = 2E^* \int_0^{a_L} [\delta - g(\tilde{a}_L)] d\tilde{a}_L = 2E^* \int_0^{a_L} u_z(x) dx\,. \tag{S9}$$

This confirms that the contact force derived from the MDR formulation agrees with the full 3D solution, validating its physical fidelity and computational utility.

Equation (S9) recovers the classical axisymmetric half-space solution within the linear reference problem. For a cylindrical flat punch of prescribed base radius $R_0$, the contact edge is fixed and the corresponding force is $F_L = 2E^*R_0\delta$.

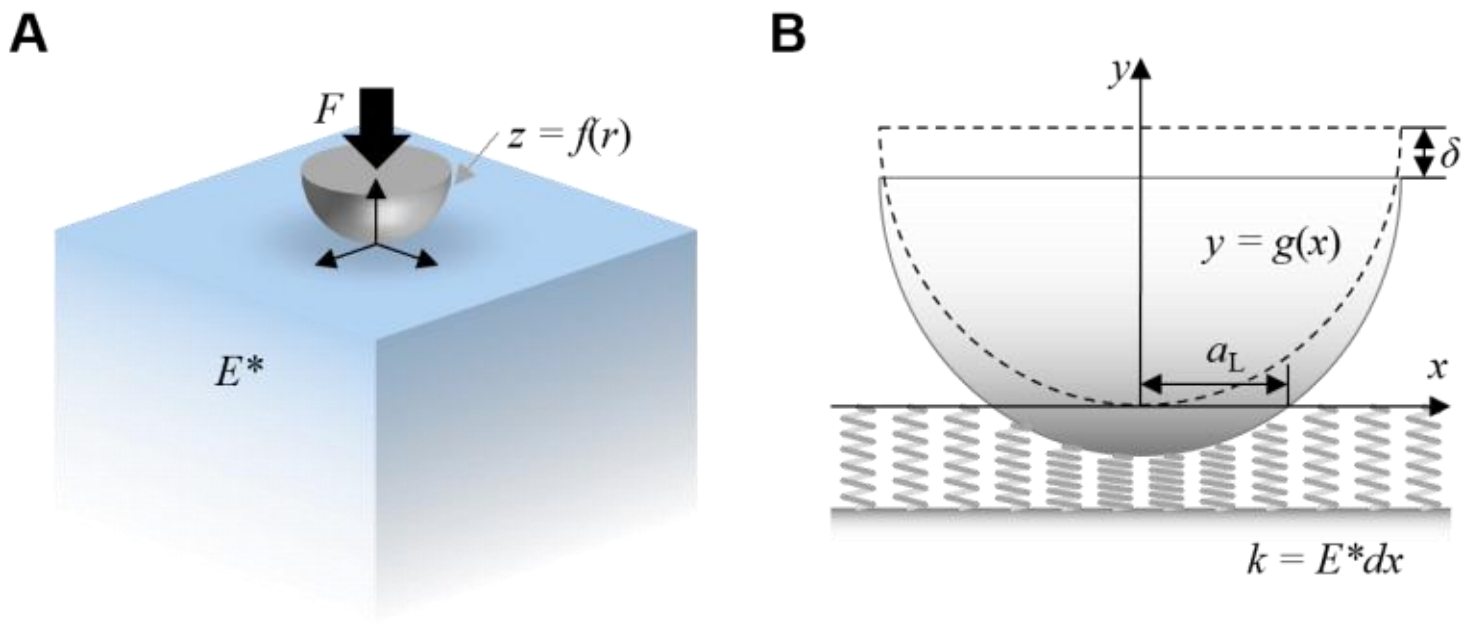

**Fig. S1. Schematics illustrating the method of dimensionality reduction (MDR) applied to an indentation problem.** (**A**) The original 3D axisymmetric contact problem involving a rigid indenter with profile $f(r)$ indenting an elastic half-space characterized by an effective modulus $E^*$. (**B**) The equivalent 1D MDR model of a rigid indenter with profile $g(x)$ on a Winkler foundation, where the stiffness of springs per length is $E^*$.

## Section S2. MDR of compression problem under large deformation

Building on the linear elasticity-based MDR framework introduced in Section S1, we extend the model to accommodate large-deformation effects in highly compliant elastomers. When the imposed compression is significant, the finite-thickness of the elastomer and associated geometric nonlinearity must be incorporated to ensure accurate force prediction.

According to Mu et al.'s analysis (9), the large-deformation compression problem can be decomposed into two parts: vertical compression and radial expansion. As shown in Fig. S2, we consider an initially axisymmetric soft structure with a curved profile $f(r)$ and initial contact radius $a_L$ (Fig. S2A). Upon compression by a vertical displacement $\delta$, the structure flattens and the contact radius expands to $a$ (Fig. S2B). To analyze this process, we employ MDR to reduce the 3D problem to an equivalent 1D model (Fig. S2C), where the elastic response is represented by a series of independent springs of spatially varying undeformed height $h(x) = L - g(x)$, and the local compression is $u_z(x) = \delta - g(x)$.

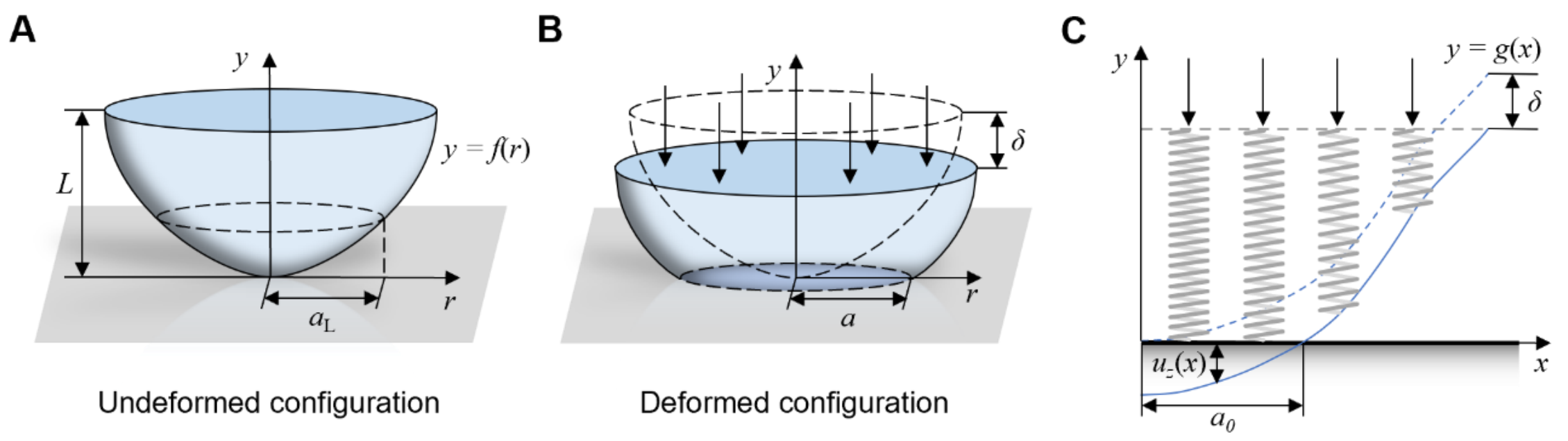

**Fig. S2. Schematics showing the model of compression problem with axisymmetric profile.** (**A** and **B**) Configurations of an axisymmetric soft body under a large-compression deformation. (**A**) Undeformed configuration of an axisymmetric soft body resting on a rigid substrate, with an initial contact radius $a_L$. (**B**) Deformed configuration of an axisymmetric soft body under compression displacement $\delta$, with the contact radius expanding to $a$ as the soft body flattens. (**C**) Corresponding 1D equivalent model based on MDR, where the 3D profile is mapped to a 1D profile $g(x)$ generated by independent springs in parallel. The compression displacement of the contacted spring located at $x$ is $u_z(x)$ under top displacement $\delta$, and the spring at the contact edge is located at $x = a_L$.

To account for geometric nonlinearity due to large deformation, Mu et al. (9) proposed a nonlinearity correction factor $\chi$, which modifies the expression of incremental contact pressure $dp_0$ caused by incremental displacement $d\delta$ of flat punch:

$$dp_0 = \chi \cdot \frac{E^*}{\pi}\frac{1}{\sqrt{a_L^2 - r^2}} d\delta, \qquad (0 \le r \le a_L). \qquad (S10)$$

For incompressible neo-Hookean materials with no profile effect, $\chi$ can be expressed as a function of normalized indentation depth $\delta/L$ (9):

$$\chi_{n.p.e.}\left(\frac{\delta}{L}\right) = \frac{1 - \frac{\delta}{L} + \frac{1}{3}\left(\frac{\delta}{L}\right)^2}{\left(1 - \frac{\delta}{L}\right)^2}. \tag{S11}$$

To incorporate the spatially varying compression due to curved geometry, the correction factor is generalized as a function of local compression ratio $\varepsilon(x) = u_z(x)/h(x)$, and then the factor $\chi$ is assumed as an average as:

$$\chi = \frac{1}{a_L}\int_0^{a_L} \chi_{n.p.e.}\big(\varepsilon(r)\big)dr = \frac{1}{a_L}\int_0^{a_L} \frac{1 - \varepsilon(r) + \frac{1}{3}\big(\varepsilon(r)\big)^2}{\big(1 - \varepsilon(r)\big)^2} dr. \tag{S12}$$

Substituting Eq. (S12) into Eq. (S10), and integrating the incremental pressure d$p_0$ on contact region, the corresponding incremental contact force is:

$$\begin{aligned} dF &= \int_0^{a_L} dp_0 2\pi r dr = 2E^* \int_0^{a_L} \left[\frac{1}{a_L}\int_0^{a_L} \chi_{n.p.e.}\big(\varepsilon(r)\big)dr\right] \cdot \frac{r}{\sqrt{a_L^2 - r^2}} dr\, d\delta \\ &= \frac{2E^*}{a_L}\left[\int_0^{a_L} \frac{r}{\sqrt{a_L^2 - r^2}} dr\right] \cdot \left[\int_0^{a_L} \chi_{n.p.e.}\big(\varepsilon(r)\big)dr\right] d\delta, \\ &= 2E^* \int_0^{a_L} \chi_{n.p.e.}\big(\varepsilon(r)\big)dr\, d\delta, \end{aligned} \tag{S13}$$

i.e., contact stiffness:

$$\frac{dF}{d\delta} = 2E^* \int_0^{a_L} \chi_{n.p.e.}\big(\varepsilon(r)\big)dr. \tag{S14}$$

Similar to Eq. (S3), the contact force is expressed as:

$$F = \int_0^{F} d\tilde{F} = \int_0^{a_L} \frac{d\tilde{F}}{d\tilde{\delta}} \frac{d\tilde{\delta}}{d\tilde{a}_L} d\tilde{a}_L = \int_0^{a_L} \frac{d\tilde{F}}{d\tilde{\delta}} g'(\tilde{a}_L) d\tilde{a}_L \tag{S15}$$

Integrating Eq. (S15) by parts, the expression of $F$ can be expressed as:

$$F = 2E^* g(a_L) \int_0^{a_L} \chi_{n.p.e.}\big(\varepsilon(r)\big)dr - \int_0^{a_L} g(\tilde{a}_L) \frac{d}{d\tilde{a}_L}\frac{d\tilde{F}}{d\tilde{\delta}} d\tilde{a}_L. \tag{S16}$$

Substituting Eq. (S2) into Eq. (S16) and simplifying, the contact force expression becomes:

$$\begin{aligned} F &= 2E^*\delta \int_0^{a_L} \chi_{n.p.e.}\big(\varepsilon(r)\big)dr - \int_0^{a_L} 2E^* \chi_{n.p.e.}\big(\varepsilon(r)\big) g(r) dr \\ &= 2E^* \left[\int_0^{a_L} \chi_{n.p.e.}\big(\varepsilon(r)\big)\delta dr - \int_0^{a_L} \chi_{n.p.e.}\big(\varepsilon(r)\big) g(r) dr\right] \\ &= 2E^* \int_0^{a_L} \chi_{n.p.e.}\big(\varepsilon(r)\big) \cdot [\delta - g(r)] dr. \end{aligned} \tag{S17}$$

Finally, noting that $u_z(x) = \delta - g(x)$, the total contact force becomes:

$$F = 2E^* \int_0^{a_L} \chi_{n.p.e.}\big(\varepsilon(x)\big) u_z(x) dx \,. \tag{S18}$$

Eq. (S18) establishes the mechanical equivalence between the original 3D compression problem and the reduced 1D MDR model with nonlinearity corrections. Dividing Eq. (S16) by the reference force in Eq. (S15) and using $n = 1 + 1/p$ yields the correction function reported in main-text:

$$F = \frac{1 - \frac{2n}{1+n}\frac{\delta}{L} + \frac{n}{2+n}\left(\frac{\delta}{L}\right)^2}{\left(1 - \frac{\delta}{L}\right)^2} F_L \tag{S19}$$

Due to the lateral expansion induced by finite deformation, the contact radius should be further corrected by accounting for radial stretch:

$$a = \int_0^{a_\mathrm{L}} \sqrt{\frac{J}{1 - \varepsilon(x)}} dx \,, \tag{S20}$$

where $J$ denotes the relative volume change. For the incompressible case $J = 1$, the same power-law substitution gives the normalized contact radius in terms of the lower incomplete beta function:

$$a = \frac{1}{p} B_{\delta/L}\left(\frac{1}{p}, \frac{3}{2}\right) \cdot \frac{(\delta/L)^{-1/p}}{\sqrt{1 - \delta/L}} \cdot a_\mathrm{L}, \tag{S21}$$

Solutions of contact radius for typical power-law profiles are listed in Table S1.

**Table S1.** Contact radius for typical power-law profiles.

| Profile, $f(r)$ | Contact radius, $a$ |
|---|---|
| $f(r) = c\,r$ | $a = \frac{2}{3}\left(1 + \frac{1/\sqrt{1-\delta/L} - 1}{\delta/L}\right) a_L, \quad a_L = \frac{2}{\pi}\frac{\delta}{c}$ |
| $f(r) = c\,r^2$ | $a = \frac{1}{2}\left(1 + \frac{\arcsin\sqrt{\delta/L}}{\sqrt{\delta/L - (\delta/L)^2}}\right) a_L, \quad a_L = \sqrt{\frac{\delta}{c}}$ |
| $f(r) = c\left(\frac{r}{R_0}\right)^p, (p \to \infty)$ | $a = \frac{1}{\sqrt{1 - \delta/L}} R_0$ |

## Section S3. Effect of constitutive model on compression response

To evaluate the influence of different constitutive models on the contact force response under large deformation, finite element analysis (FEA) simulations were performed using three representative hyperelastic models: neo-Hookean (10), Mooney-Rivlin (11, 12), and Arruda-Boyce (13, 14). All models were calibrated to share the same initial shear modulus ($\mu$ = 1 MPa) and Poisson's ratio ($\nu$ = 0.5), thereby ensuring consistent behavior in the small-strain linear regime (Fig. S3A).

The Arruda–Boyce model exhibits pronounced strain stiffening due to its basis in molecular chain network theory, which accounts for finite chain extensibility. As the locking stretch parameter $\lambda_m$ approaches infinity, the model asymptotically reduces to the neo-Hookean form. Typical values of $\lambda_m$ range from 4 to 6 for elastomeric materials such as rubbers (13, 14), SMPs (15-17), and hydrogels (18, 19).

For the Mooney–Rivlin model, the shear modulus is given by $\mu = 2(C_{10} + C_{01})$. A common choice for rubber-like elastomers is $C_{01} \ll C_{10}$ (14), and accordingly, simulations were performed using $C_{10}$ = 0.45 and 0.40 with $C_{01}$ values determined by shear modulus $\mu$.

Figures S3B-D show the computed contact force versus normalized displacement for the compression of three canonical geometries (cone, hemisphere, and cylinder) under these different constitutive models. Despite inherent differences in their nonlinear stress-strain behavior, all models yield remarkably similar force responses across the geometries examined, as long as their linear elastic moduli are matched. These results indicate that, within the deformation ranges considered ($\delta/L \leq 0.8$), the choice of hyperelastic model has only a marginal effect on the global contact response.

Therefore, the neo-Hookean model is adopted in the main analysis. Its simplicity facilitates analytical treatment while still capturing the essential features of hyperelastic material behavior, making it a practical and sufficiently accurate choice for the problem studied.

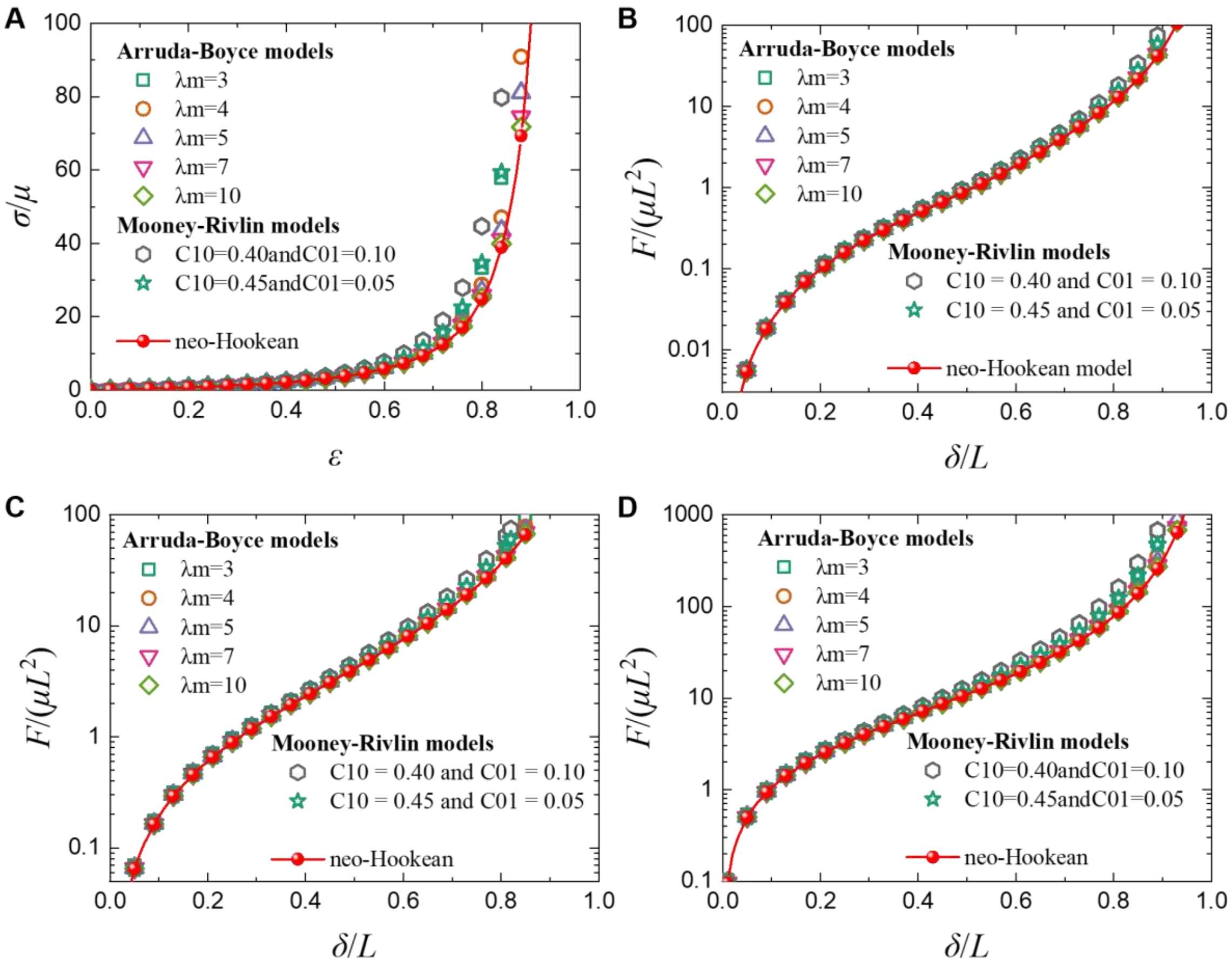

**Fig. S3. Comparison of different constitutive models and their influence on the contact force response during compression of representative geometries, as obtained from FEA.** (**A**) Uniaxial stress–strain curves under simple compression for the Arruda-Boyce, Mooney-Rivlin, and neo-Hookean models, all with identical shear modulus ($\mu$ = 1 MPa) and Poisson's ratio ($v$ = 0.5) in the linear regime. (**B**, **C**, and **D**) Contact force versus normalized compression displacement for (**B**) conical, (**C**) hemispherical, and (**D**) cylindrical profiles, each modeled using the corresponding hyperelastic material models.

To examine the role of material compressibility, finite element simulations were conducted using neo-Hookean materials with identical shear modulus but different Poisson's ratios. The shear modulus $\mu$ of the linear elastic model is fixed at 1 MPa, and Poisson's ratio $v$ varies from 0.3 to 0.5, covering the typical range of common soft elastomers (9, 20, 21). Figure S4 shows the reciprocal correction functions for cone, sphere, and cylinder geometries.

The results indicate that variations in Poisson's ratio produce only minor changes in the correction functions, particularly at moderate and large compression. The responses for different Poisson's ratios largely collapse onto a single trend, demonstrating that the nonlinear compression behavior is dominated primarily by geometric effects rather than material compressibility. This weak dependence is consistent across all geometries considered.

These results justify the use of a single correction function primarily governed by geometry and shear modulus in the proposed framework, and support the applicability of the analytical model to a broad class of soft materials without requiring precise knowledge of Poisson's ratio.

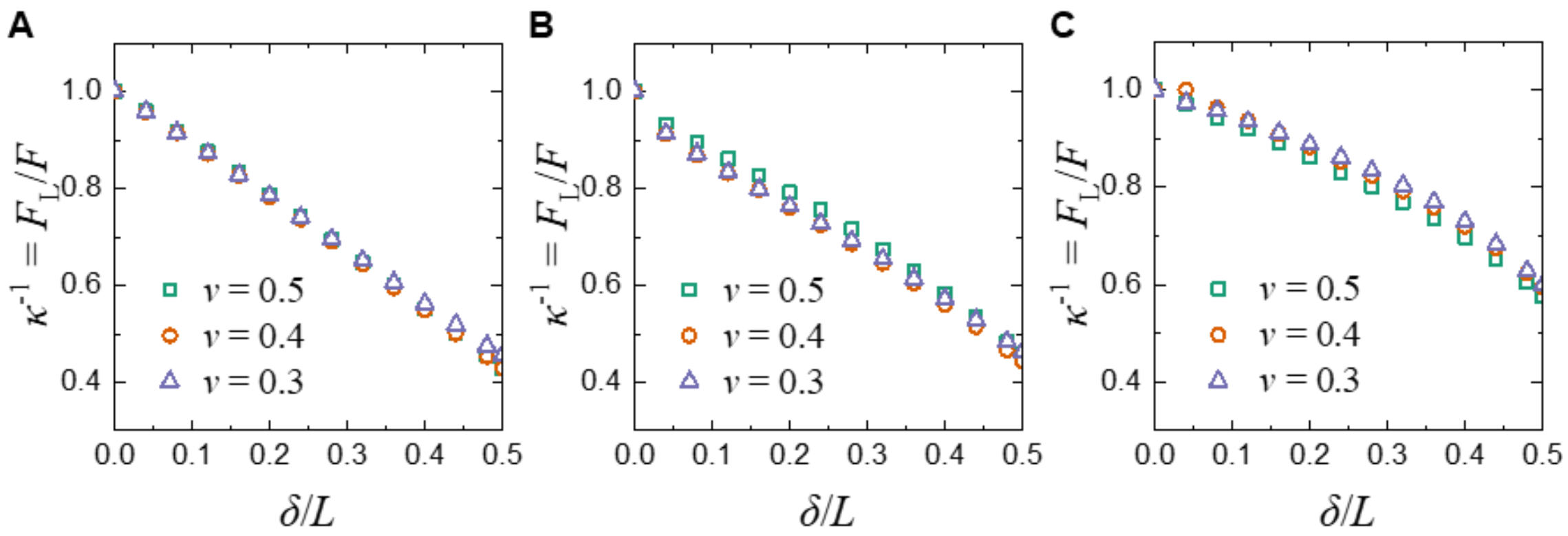

**Fig. S4. Influence of Poisson's ratio under extreme compression.** Reciprocal correction functions obtained from FEA for (**A**) cone, (**B**) sphere, and (**C**) cylinder geometries. All simulations employ neo-Hookean material models with an identical shear modulus ($\mu$ = 1 MPa), while the Poisson's ratio is varied to evaluate its effect on the large-deformation contact response.

## Section S4. Linear solution of contact force for power-law profiles

For an axisymmetric power-law profile $f(r) = c\, r^p$, the contact force under linear elasticity is given by Eq. (S9) as (1, 2, 5, 22-24):

$$F_L = C \cdot \delta^{1+\frac{1}{p}}, \tag{S22}$$

where $\delta$ is the indentation depth, and $C = E^* \frac{2p}{1+p}[cK(p)]^{-\frac{1}{p}}$. Here, $E^*$ is the effective modulus, and $K(p) = \sqrt{\pi}\,\Gamma(p/2+1)/\Gamma(p/2+1/2)$.

The contact radius is given by Eq. (S2) as:

$$a_L = D \cdot \delta^{\frac{1}{p}}, \tag{S23}$$

where $D = [cK(p)]^{-\frac{1}{p}}$.

For a hemisphere of radius $L$, the surface near the contact region can be approximated by a parabolic profile, i.e., $f(r) = r^2/(2L)$. In this case, $c = 1/(2L)$, and the parameter $C = 4/3\ E^*\ L^{1/2}$.

For a cone, this reduces to $c = \tan\theta$ and $p = 1$, yielding the classical force-displacement relation from Eqs. (S22) as:

$$F_{L,0} = \frac{2E^*}{\pi \tan\theta}\delta^2. \tag{S24}$$

Similarly, the corresponding classical expression for contact radius is obtained from Eqs. (S23) as:

$$a_{L,0} = \frac{2\cot\theta}{\pi}\delta. \tag{S25}$$

However, these solutions assume an infinite half-space and neglect geometric constraints of the cone. As shown in Figs. S5A-B, when the cone angle $\theta$ is small, the available elastic volume near the tip becomes significantly limited, reducing the effective elastic response. This discrepancy is visualized through FEA simulations (Fig. S5B), where von Mises stress distributions show reduced support beneath the contact interface at finite deformation ($\delta/L = 0.01$).

Figures S5C-D compare FEA predictions (blue circles) with the classical linear solutions (gray dashed lines), which consistently overestimate the contact force and underestimate the contact radius. To account for the geometric confinement, we introduce angle-dependent correction factors $A_F(\theta)$ and $A_a(\theta)$, modifying the classical predictions (Eqs. (S24) and (S25)) as:

$$F_L = A_F(\theta) \cdot F_{L,0}, \qquad a_L = A_a(\theta) \cdot a_{L,0}. \tag{S26}$$

where corrections enable accurate modeling across a wide range of cone angles.

Fig. S5E summarizes the extracted correction factors from FEA. The red symbols and line represent $A_F(\theta)$ for contact force, while the black symbols and line represent $A_a(\theta)$ for contact radius. The fitting equations are given as:

$$A_F(\theta) = 1 - \left(1 - \frac{2\theta}{\pi}\right)^{2.5}, \qquad A_a(\theta) = 1 - 0.0018 \times \left(\theta - \frac{\pi}{2}\right). \tag{S27}$$

Both factors exhibit strong dependence on $\theta$, confirming the geometric origin of the deviation from the classical solution. The corrected model aligns well with numerical results and provides a refined framework for conical compression problems.

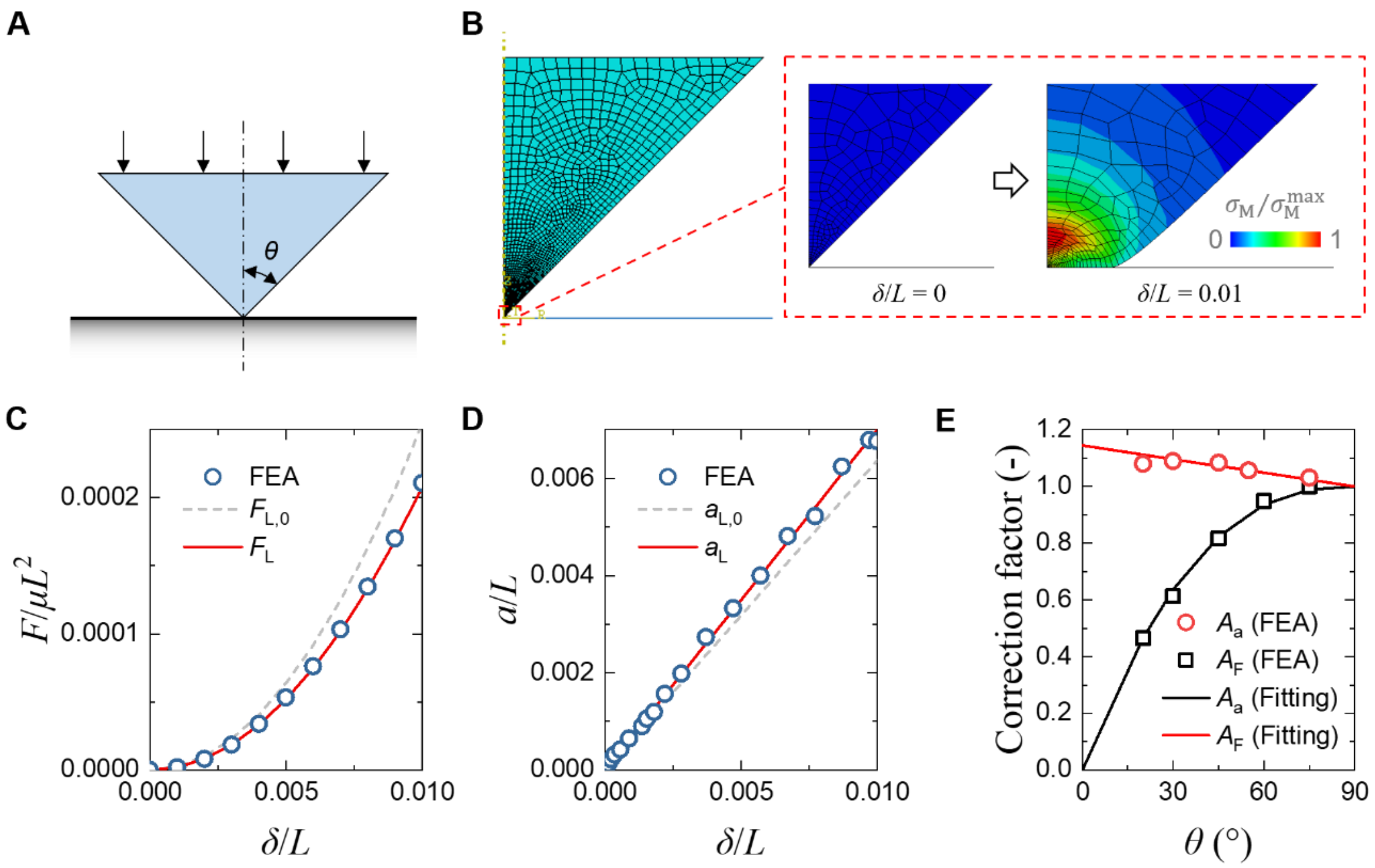

**Fig. S5. Compression analysis of an elastic cone.** (**A**) Schematic of contact between an elastic cone with cone angle $\theta$ and a rigid substrate. (**B**) FEA model using a cone with $\theta = 45°$ as an example. The undeformed geometry is shown on the left, while insets display von Mises stress distributions near the contact region in the undeformed ($\delta/L$ = 0) and compressed ($\delta/L$ = 0.01) states. (**C**) Contact force and (**D**) and contact radius as functions of normalized displacement. Blue circles represent FEA results; gray dashed lines denote classical linear solutions; red solid lines show modified linear solutions incorporating cone angle correction. (**E**) Correction factors for contact force (red circles and line) and contact radius (black squares and line) as functions of cone angle.

## SI References